\documentclass[twocolumn,showpacs,preprintnumbers,amsmath,amssymb,floatfix]{revtex4}

\usepackage{graphicx}
\usepackage{dcolumn}
\usepackage{bm}

\newcommand{\beq}{\begin{equation}}
\newcommand{\eeq}{\end{equation}}
\newcommand{\bey}{\begin{eqnarray}}
\newcommand{\eey}{\end{eqnarray}}

\newcommand{\kpc}{\, {\rm kpc} }

\newcommand{\Msun}{M_\odot} 
 
\newcommand{\vs}{{\bf s}}
\newcommand{\vg}{{\bf g}}

\newcommand{\vx}{{\bf x}}
\newcommand{\vD}{{\bf D}}
\newcommand{\vE}{{\bf E}}
\newcommand{\vP}{{\bf P}}
\newcommand{\vd}{{\bf d}}

\newcommand{\vp}{{\bf p}}
\newcommand{\grad}{{\bf \nabla}}

\newcommand{\mut}{{\tilde\mu}}

\begin{document}

\preprint{APS/123-QED}

\title{Insight into the baryon-gravity relation in galaxies}

\author{Benoit Famaey} 
\email{bfamaey@ulb.ac.be}
\affiliation{Institut d'Astronomie et d'Astrophysique, Universit\'e Libre de Bruxelles, Boulevard du Triomphe CP226, B-1050, Bruxelles, BELGIUM}%

\author{Gianfranco Gentile} \affiliation{University of New Mexico, Department of Physics and Astronomy, 800 Yale Blvd NE, Albuquerque, New Mexico 87131, USA}%

\author{Jean-Philippe Bruneton} \affiliation{$\mathcal{G}\mathbb{R}\varepsilon\mathbb{C}\mathcal{O}$, Institut d'Astrophysique de Paris, UMR 7095-CNRS, Universit\'e Pierre et Marie Curie - Paris 6, 98 bis boulevard Arago F-75014, Paris, FRANCE}%

\author{HongSheng Zhao} \affiliation{SUPA, School of Physics and Astronomy, University of St Andrews, KY16 9SS, Fife, UK}%

\date{\today}

\begin{abstract}
Observations of spiral galaxies strongly support a one-to-one analytical relation between the inferred gravity of dark matter at any radius and the enclosed baryonic mass. It is baffling that baryons manage to settle the dark matter gravitational potential in such a precise way, leaving no ``messy" fingerprints of the merging events and ``gastrophysical" feedbacks expected in the history of a galaxy in a concordance Universe. This correlation of gravity with baryonic mass can be interpreted from several non-standard angles, especially as a modification of gravity called T$e$V$e$S, in which no galactic dark matter is needed. In this theory, the baryon-gravity relation is captured by the dieletric-like function $\mu$ of Modified Newtonian Dynamics (MOND), controlling the transition from $1/r^2$ attraction in the strong gravity regime to $1/r$ attraction in the weak regime. Here, we study this $\mu$-function in detail. We investigate the observational constraints upon it from fitting galaxy rotation curves, unveiling the degeneracy between the stellar mass-to-light ratio and the $\mu$-function as well as the importance of the sharpness of transition from the strong to weak gravity regimes. We also numerically address the effects of non-spherical baryon geometry in the framework of non-linear T$e$V$e$S, and exhaustively examine how the $\mu$-function connects with the free function of that theory. In that regard, we exhibit the subtle effects and wide implications of renormalizing the gravitational constant. We finally present a discontinuity-free transition between quasi-static galaxies and the evolving Universe for the free function of T$e$V$e$S, inevitably leading to a return to $1/r^2$ attraction at very low accelerations in isolated galaxies.
\end{abstract}

\maketitle

\section{Introduction}

Data on large scale structures point towards a Universe dominated by dark matter (DM) and dark energy \cite{Spergel, Clowe}. Nowadays, the dominant paradigm is that DM is actually made of non-baryonic weakly-interacting massive particles, the so-called {\it cold dark matter} (CDM). However, on {\it galaxy scales}, the observations appear to be at variance with a sizeable list of CDM predictions. Measurements of non-circular motions in the
Milky Way have shown that there is actually very little room for an axisymmetric distribution of dark matter inside the solar radius \cite{Bis03, FB05}, where CDM simulations predict a cuspy density profile \cite{Di}. Even though CDM halos themselves could well be triaxial \cite{triax}, the non-circular motion of gas in the inner Milky Way is proven to be linked with non-axisymmetric {\it baryonic} features such as the bar and the spiral arms \cite{Bis03}, thus leaving no room for a cuspy triaxial CDM halo to explain them. External galaxies have also been used to compare the predicted cuspy CDM density profiles with the observations, in particular rotation curves of dwarf and spiral galaxies. Despite the presence
of observational systematic effects \cite{Swa}, most of which are now
under control, observations point towards dark matter halos with a
central constant density core \cite{deblok, Gen1, Gen2, Simon, Kuzio} at odds with the CDM predictions. Another interesting problem faced by CDM on galactic scales is the overabundance of predicted satellite galaxies compared
to the observed number in Milky Way-sized galaxies \cite{Moore}. 

What is more, it is now well-documented that rotation curves suggest a {\it correlation} between the mass
profiles of the baryonic matter (stars + gas) and dark matter \cite{Don, McG, McGa}. Some rotation curves even display obvious features (bumps  or wiggles) that  are also clearly  visible in  the stellar  or gas
distribution \cite{Broeils}. A solution to all these problems, and especially the baryon-DM relation, could be some kind of new specific interaction between baryons and some exotic DM \cite{McG}. On the other hand, it could
indicate that, on galaxy scales, the observed
discrepancy rather reflects a breakdown of Newtonian dynamics in the
ultra-weak field regime: this original proposal, dubbed MOND (Modified Newtonian Dynamics), was made more than twenty years ago in a series of hallmark papers \cite{Milgrom83, Milg2, Milg3}. But as noted in, e.g., \cite{Milgrom05}, a modified gravity theory is
in many ways {\it equivalent} to an exact relation between baryons and dark
matter. When analyzing rotation curves, the two interpretations are
nearly degenerate in the sense that the fraction of
gravity unaccounted by the Newtonian gravity of the baryons at radius $R$ is
\beq
1 - \frac{\xi
G_N M_{b}(R)}{R^2 g} = \begin{cases} \frac{G_N M_{DM}(R)}{R^2 g } & {\rm in} \,
{\rm DM} \, {\rm picture} \cr   \left[1 - \mu(g/a_0) \right] & {\rm in} \,
{\rm MOND} \, {\rm picture} \cr
\end{cases}
\eeq
where $G_N$ is the usual Newtonian gravitational constant, $M_b(R)$ is the baryonic mass interior to the galactocentric radius $R$, $\frac{\xi G_N M_b(R)}{R^2}$ denotes the Newtonian gravity of the
baryons (with a shape factor $\xi \sim 1$  to correct for
non-spherical baryon distribution), and  $g =
v^2/R$ is the total centripetal gravitational acceleration
from measuring the circular speed $v$ at radius $R$. The first equality introduces a
spherical halo of dark matter with mass  $M_{\rm DM}(R)$ enclosed inside
radius $R$ (a flattening factor can also be used for the DM halo).  The second equality (Milgrom's law) invokes a modified dynamics with a
dielectric-like \cite{Blanchet1, Blanchet2} factor $\mu \le 1$ to boost Newtonian gravity below
a dividing acceleration scale $a_0$ whose value turns out to be of the same order as $cH_0$. This $\mu$ factor is also called the MOND interpolating function $\mu(x)$ (where $x=g/a_0$).

The {\it key difference} between the usual dark matter and the MOND interpretation is the fact that MOND requires all galaxies to share the same $\mu$-function for the same value of $a_0$. If it is not the case, then the present-day success of MOND just tells us something phenomenological about DM. On the other hand, if it is indeed exactly the case, then either a to-be-discovered exotic relation does actually exist between DM (not the usual CDM) and baryons, or the DM paradigm is incorrect on galaxy scales. In the latter case, either gravity should be modified \cite{BM84, Bekenstein2004}, which could lead to some subtle differences regarding e.g. dynamical friction \cite{dynf1, dynf} or tides \cite{Zglob, ZT}, or inertia should be modified \cite{Milgrom05, Milgrom94, Milgrom99}, which would lead to very significant differences with the DM interpretation, especially in non-stationary situations \cite{Milgrom05}.

The basis of the MOND paradigm \cite{Milgrom83} is that $\mu$ is a function which runs smoothly from $\mu(x)=1$ at $x\gg 1$ to $\mu(x)=x$ at $x\ll 1$. Even though this leaves some freedom for the exact shape of $\mu$, it is already extremely surprising, from the DM point of view, that such a prescription did predict many aspects of the dynamics of galaxies such as the appearance of large discrepancies with Newtonian dynamics in low surface brightness galaxies, as well as the absence of discrepancies in the inner parts of high surface brightness ones. With this definition of the $\mu$-function, the fraction of gravity unaccounted by Newtonian gravity tends to zero in the high-acceleration limit, and one recovers in the low-acceleration limit the Tully-Fisher law \cite{Tully, McG2}, which tells us that the circular velocity at the last measured radius in spiral galaxies must behave like $M_b^{1/4}$. Since the centripetal acceleration at that radius is typically lower than $\sim 0.4 a_0$ in high surface brightness galaxies (and even lower in low surface brightness galaxies), the MOND circular velocity at that radius is within at most ten percent of the asymptotic one, depending on the exact shape of the $\mu$-function, and on the stellar mass-to-light ratio. Indeed, the exact shape of the $\mu$-function is still somewhat ambiguous in MOND, because there is a degeneracy between the choice of $\mu$ and the stellar mass-to-light ratio when rotation curves are fitted with Eq.~(1). While the exact asymptotic velocity will actually be different depending on the mass-to-light ratio, the shape of the $\mu$-function will at the same time determine at what radius the rotation curve {\it really} becomes flat, and this can happen somewhat away from the last observed radius, thus explaining this degeneracy.

Even though the most important aspect of MOND, confronting CDM with the most severe observational difficulties, remains its asymptotic behaviour described above, the study of the transition between the Newtonian regime and the asymptotic MOND regime, characterized by the shape of the $\mu$-function, may also be important in the quest for a possible fundamental theory underpinning the MOND prescription. In this contribution, we thus investigate the observational constraints upon the $\mu$-function in that transition zone, as well as the theoretical constraints upon it in the framework of Bekenstein's relativistic multifield theory of gravity \cite{Bekenstein2004}.

\section{Observational constraints on the $\mu$-function}

\subsection{The $\alpha$-family and the standard function}

In this section, we examine the observational constraints upon the $\mu$-function from fitting galaxy rotation curves with Milgrom's law (the second equality of Eq.~1). As a matter of fact, we shall prove in Sect.~III~F that this empirical law yields an excellent approximation to Bekenstein's relativistic multifield theory of MOND \cite{Bekenstein2004} in fitting rotation curves of spiral galaxies. 

Here, we explicitly show the degeneracy between the stellar mass-to-light ratio (+ the distance of the galaxy) and the shape of the $\mu$-function, a degeneracy that prevents from fixing the $\mu$-function in present-day MOND. Among  a parametric family of $\mu$-functions, we are going to show that the best fits are obtained for the ``simple" $\mu$-function of \cite{FB05}:
\beq \protect\label{simple}
\mu(x) =\frac{x}{1+x},
\eeq
and that, notwithstanding the $M_*/L$-$\mu$ degeneracy (where $M_*/L$ is the stellar mass-to-light ratio), $\mu$-functions with too slow a transition from the Newtonian to the deep-MOND regime are excluded by rotation curves data, while the ``standard" $\mu$-function \cite{SanMc}, \beq \protect\label{standard} \mu(x)= \frac{x}{\sqrt{1+x^2}}, \eeq
with a faster transition than Eq.~(\ref{simple}), yield fits of comparable quality.

The set of $\mu$-functions we propose to test against observations is called the $\alpha$-family \cite{AFZ}:
\beq
\mu(x) = \frac{2x}{1+ (2-\alpha) x + \sqrt{(1-\alpha x)^2 + 4x}}, ~ 0
\leq \alpha \leq 1.
\label{eqalpha}
\eeq
In the $\alpha=1$ case, one directly
recovers the simple function of Eq.~(\ref{simple}). In the
$\alpha=0$ case, it is a simple exercise to show that one recovers the $\mu$-function of Bekenstein's toy-model \cite[eq.(62)]{Bekenstein2004}
\beq \protect\label{bekmu}
\mu(x) = \frac{-1 + \sqrt{1+4x}}{1+\sqrt{1+4x}}.
\eeq

\begin{figure}
   \centering
   \includegraphics[width=8cm]{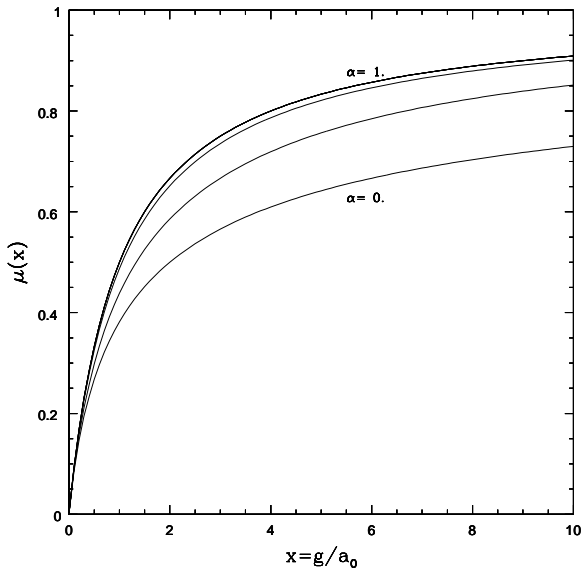}
      \caption{The $\mu$-functions of Eq.~(\ref{eqalpha}) for $\alpha=0$, $\alpha=0.5$, $\alpha=0.9$, and $\alpha=1$, illsutrating that the transition from the deep-MOND ($\mu \sim x$) to the Newtonian ($\mu \sim 1$) regime is sharper with increasing $\alpha$.
\label{alphamu}
       }
   \end{figure}

Fig.~1 displays those $\mu$ as a function of $x$. This is of course far from testing all the other possibilities, notably the free function proposed in \cite{sanderssolar}, but it will give a good indication of the range of allowed $\mu$-functions to be compatible with observed rotation curves. 

\subsection{Fitting galaxy rotation curves}

In order to test observationally the parametric $\alpha$-family of
$\mu$-functions, we selected a sample made of two subsamples of
rotation curves: the first subsample is made of nine rotation curves presented
in \cite{BBS}. This sample has been used to determine
the value of the acceleration constant $a_0=1.21 \times 10^{-8}$ cm s$^{-2}$ that is
generally used in MOND studies with the standard $\mu$-function. The second subsample consists of the five
galaxies presented in \cite{Gen1}. This way, we have a final
sample of 14 galaxies with  high-quality, extended rotation curves. These
galaxies span a large range of Hubble types and maximal velocities, from
small irregular dwarf galaxies to large early-type spirals.

The values of $a_0$ and $\alpha$ are unknown a priori, but have to be the
same for all the galaxies. Therefore, we performed
a global fit for these two parameters to the abovementioned 14 rotation curves:
$\alpha$ was allowed to vary from $0$ to $1$, while the stellar $M/L$ ratio and the
distance ($\pm30\%$ the distances adopted in \cite{BBS} and \cite{Gen1}) were left as parameters free to vary individually for
each galaxy (see Table I).

\begin{table}
\centering
\caption{Best-fit distances ($d$ in Mpc) and stellar $M/L$ ratios (in solar units) of the MOND rotation curve fits with the standard $\mu$-function (Eq.~\ref{standard}) and the simple $\mu$-function (Eq.~\ref{simple}, $\alpha=1$). Distances were left as free parameters within 30\% of the distances adopted in the original papers; for NGC 7331 the bulge $M/L$ was considered as an additional free
parameter, with a best-fit value of 1.5 for both $\mu$-functions. The degeneracy between those free parameters and the choice of the $\mu$-function appears clearly ($M_*/L$ playing the most important role in this regard).}
\label{}
\begin{tabular} {c c c c c c c c c c c c c c c l c c r}
\hline
  & & & & & & & & & & & & & & & Standard $\mu$ & & & Simple $\mu$  \\
\end{tabular}
\begin{tabular} {l c c c c}
\hline
Galaxy & $d$ & $M_*/L$ & $d$ & $M_*/L$ \\
\hline
NGC 2403 & 3.5 & 1.2 & 3.3 & 0.9 \\
NGC 2903 & 6.5 & 3.5 & 5.1 & 3.4 \\
NGC 3198 & 7.2 & 4.1 & 7.1 & 2.9 \\
NGC 6503 & 5.5 & 2.0 & 5.5 & 1.4 \\
NGC 7331 & 13.3 & 5.0 & 13.6 & 2.4 \\
NGC 1560 & 3.4 & 0.4 & 3.0 & 0.4 \\
NGC 2259 & 9.5 & 2.3 & 9.4 & 1.6 \\
DDO 154 & 3.4 & 0.1 & 3.0 & 0.1 \\
DDO 170 & 14.0 & 0.7 & 12.7 & 0.7 \\
ESO 116-G12 & 20.0 & 0.3 & 19.8 & 0.2 \\
ESO 287-G13 & 34.8 & 1.4 & 31.1 & 1.1 \\
ESO 79-G14 & 28.1 & 2.1 & 25.6 & 1.7 \\
NGC 1090 & 25.5 & 2.3 & 25.5 & 1.5 \\
NGC 7339 & 23.4 & 1.4 & 23.4 & 0.9 \\
\hline
\end{tabular}
\end{table}

The best fits are obtained for
\beq
\alpha=1^{+0.0}_{-0.5},
\eeq
and
\beq\protect\label{simplea0}
a_0=1.35^{+0.28}_{-0.42} \times 10^{-8} {\rm cm} \, {\rm s}^{-2},
\eeq
where the errors are the one-sigma uncertainties from the $\chi^2$ statistics of the global fit; this value of $a_0$ is compatible with the result of \cite{BBS}. For the unfavoured functions with $\alpha<0.5$, the transition from the deep-MOND to the Newtonian regime is slower than for $\alpha>0.5$ as shown on Fig.~1. 

\begin{figure*}
   \centering
   \includegraphics[width=16cm]{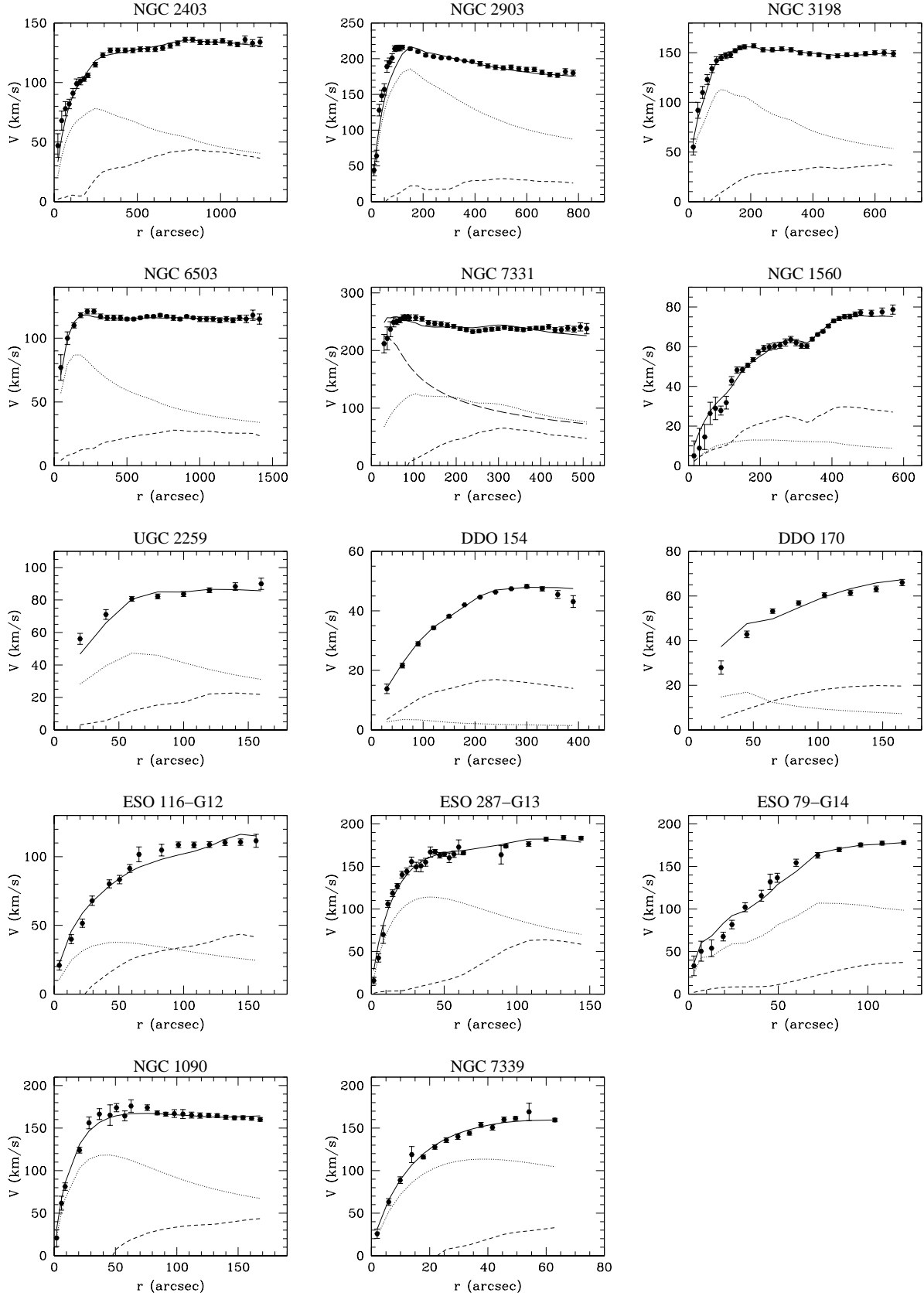}
      \caption{
Rotation curve decompositions with $\alpha=1$ [see Eq.(\ref{eqalpha}) and Eq.(\ref{simple})]
and $a_0=1.35 \times 10^{-8}$ cm s$^{-2}$. The distances were left as free
parameters, within 30\% of the adopted distances in the original papers.
The solid line is the best fit to the data,
the dotted/dashed/long-dashed lines represent the disk/gas/bulge Newtonian contributions. Note that the Newtonian curves deviate from the MOND ones even at small radii, which is characteristic of the simple function $\alpha=1$.
\label{rc}
       }
\end{figure*}

\begin{figure*}
   \centering
   \includegraphics[width=16cm]{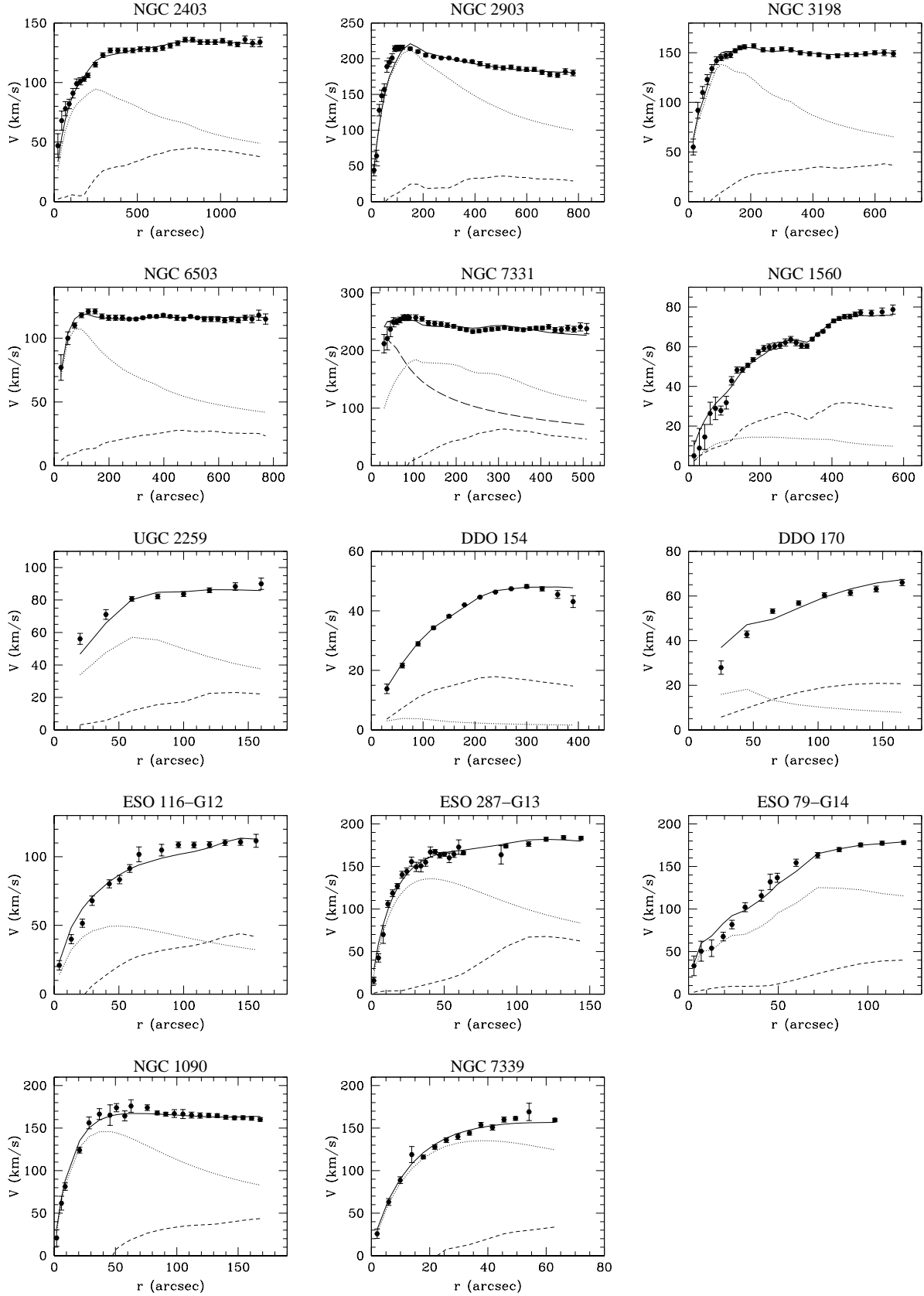}
      \caption{
Rotation curve decompositions with the standard MOND $\mu$-function [see Eq.(\ref{standard})], and $a_0=1.21 \times 10^{-8}$ cm
s$^{-2}$. The distances were left as free parameters, within 30\% of the
adopted distances in the original papers. The solid line is the best fit to
the data, the dotted/dashed/long-dashed lines represent the disk/gas/bulge Newtonian contributions. Note that the distances were already allowed to vary in fits with the standard function in \cite{BBS} (and that the fits presented on this figure are thus the same) but not in \cite{Gen1}.
\label{rc_old}
       }
\end{figure*}

The results of the rotation curve fits
using the best fit parameters $\alpha=1$ and $a_0=1.35\times 10^{-8}$ cm
s$^{-2}$ are shown in Fig. \ref{rc}. When comparing the fits
of Fig. \ref{rc} to those performed with the standard MOND $\mu$-function (Eq.\ref{standard}, Fig. \ref{rc_old}), we see that the simple $\mu$-function clearly
fits the rotation curves as well as the standard one: the $\chi^2$ values
of the fits are very similar. On average, we have:
\beq \label{chi}
\chi^2_{\alpha=1}/\chi^2_{standard} = 0.98\pm0.24.
\eeq
Note however that three galaxies (NGC~2403, NGC~3198, and NGC~7331) 
also have Cepheid-based distances \cite{Bottema}. While for 
NGC~2403 
and NGC~7331 the best-fit distances (of both $\mu$-functions) are within 
10\% of the Cepheid distances, the best-fit distance of NGC~3198 
is about a factor 2 smaller than the Cepheid-based one. Imposing this 
value of the distance slightly decreases the best-fit value of $a_0$ 
but it doesn't change the best-fit value of $\alpha$, nor does it change 
significantly the  result of Eq.~(\ref{chi}), because the fits with both $\mu$ functions get worse for NGC~3198. 

The relatively large error bars on $\alpha$ are due
to the degeneracy between the shape of $\mu$ and the stellar $M/L$
ratio. Still, it is striking that, notwithstanding this degeneracy, $\alpha=1$, corresponding to the
simple relation of Eq.~(\ref{simple}), actually yields the best fit. This function was also shown to yield an excellent fit to the Terminal Velocity Curve of the Milky Way Galaxy \cite{FB05}, and was found to work better than the standard function to fit the dispersion profile of the globular cluster system of NGC~4636 in the transition regime between the Newtonian and the deep-MOND regime \cite{Schu}.
Even more
interesting is to check that the stellar $M/L$ ratios implied by this simple $\mu$-function are realistic. On average, we have:
\beq
\frac{(M_*/L)_{\alpha=1}}{(M_*/L)_{\rm standard}}=0.70\pm0.15.
\eeq
i.e. their value is $\sim$30\% lower than when using the standard MOND
interpolating function. In \cite{McG} the standard
MOND $M/L$ ratios are slightly higher than the predictions
of stellar population
synthesis models (by $\sim$ 0.10 dex, using his fits of the
stellar $M/L$ vs colour plots for B-V$>$0.55, i.e. above the break),
for a given observed colour.
Recently, \cite{dB} compared the ranges of $M/L$ ratios allowed
by different methods to determine them, and suggested that the
normalisation of  the stellar $M/L$ vs colour relation should be lowered by
0.05 to 0.10 dex. Note that they consider the predictions of the models by
\cite{Bd}, while \cite{McG} uses the most recent models by
\cite{bell}.  However, in the range of colours considered by \cite{McG}, 0.4 $\lesssim$ B-V $\lesssim$ 0.8, the two models give predictions that differ by at most 0.03 dex.
The stellar $M/L$ ratios that we find from the simple  $\mu$-function,
$\sim 30\%$ lower than the standard MOND ones,
fall perfectly in this range allowed by the various constraints
considered by \cite{dB}.

Let us end this section by noting that the ``traditional" view on this observational baryon-gravity relation would be that it is a purely empirical relation that will be understood within the framework of the concordance $\Lambda$CDM cosmological model. Even though this is a very challenging task, the meaning of the $\alpha$-family of Eq.~(\ref{eqalpha}) combined with Eq.~(1) would then correspond (in spherical symmetry) to the following empirical relation:
\begin{equation}\protect\label{empirical}
\frac{g_{Nb}}{g_{DM}} = \frac{\alpha g_{Nb} + g_{DM}}{a_0},
\end{equation}
where $g_{Nb}$ and $g_{DM}$ are the contribution to the rotation curve gravity from the baryons and DM respectively. If neither gravity nor inertia are to be modified, the present success of the simple relation $\alpha=1$ to fit galaxy rotation curves still captures the essence of the fine-tuning problem for dark matter. The mass-to-light ratios of Table I being compatible with current population synthesis models, Eq.~(\ref{empirical}) tells us that the ratio of baryonic gravity to DM gravity should be equal to the total gravity in units of $a_0$. For example, if one plots the contribution of baryons and dark matter to the rotation curve at a certain radius as a function of the baryonic contribution to gravity at that radius (Fig.~4), one finds a cross reminiscent of \cite{McG}.

\begin{figure}
   \centering
   \includegraphics[width=8cm]{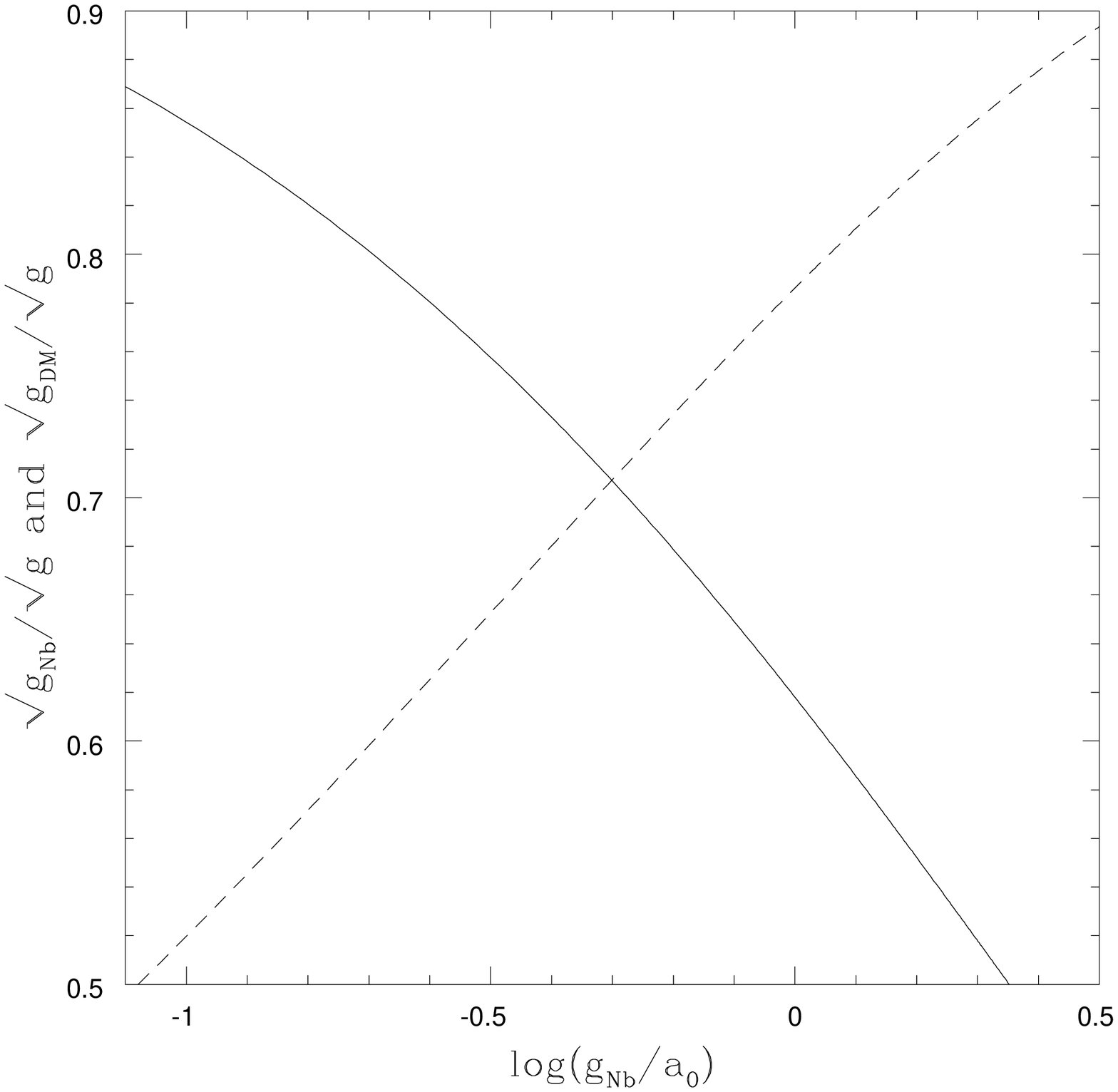}
      \caption{This figure illustrates that Eq.~(\ref{empirical}) captures the essence of the fine-tuning problem for CDM in galaxies. For $\alpha=1$, we plot the square root of the fraction of total gravity accounted by the baryons  (dashed line) and the DM (full line) as a function of the logarithm of baryonic gravity. The cross is reminiscent of the one obtained in \cite[fig.4]{McG}.
\label{dm}
       }
   \end{figure}

\section{The $\mu$-function and the free function of T$e$V$e$S}

\subsection{T$e$V$e$S in galaxies}

Here, we present the theoretical constraints upon the $\mu$-function in the framework of Bekenstein's relativistic multifield theory of gravity \cite{Bekenstein2004}. Before doing that, we need to briefly look back on the basics of the theory, and recap the relevant equations (in Sects. III~A and III~B).

If one wants to modify Newtonian gravity on galaxy scales in order to get MOND, one can modify Poisson's equation in order to recover Milgrom's law (second equality in Eq.~1) in highly symmetric systems \cite{BM84}, namely:
\beq \protect\label{eqn:poissonaqual}
\nabla.[\mu(|\grad\Phi|/a_0) \nabla \Phi ] = 4 \pi G_N \rho,
\eeq
where $G_N$ is the usual Newtonian gravitational constant, $\Phi$ is the MOND gravitational potential and $\rho$ is the density of the (baryonic) matter distribution.
In spherical symmetry, one then immediately finds back Milgrom's law (Eq.~1) from Gauss's theorem. Out of spherical symmetry, one finds that
\beq
\mu(|\grad\Phi|/a_0) \nabla \Phi = \nabla \Phi_N + \nabla \times {\bf h},
\eeq
where ${\bf h}$ is a regular vector field determined (up to a gradient) by the condition that the curl of the total force $\nabla \Phi$ must vanish. Inside spiral galaxies, a numerical MOND potential solver \cite{Brada1, CLN, Tiret} must thus be used to obtain the exact MONDian force field corresponding to Eq.~(\ref{eqn:poissonaqual}). However, it has been shown that Milgrom's law is still a good enough approximation to fit galactic rotation curves within that framework \cite{Brada1, Brada2}.

On the other hand, the extension of General Relativity that has been proposed by Bekenstein for the MOND paradigm is a {\it bi-metric multi-field} theory \cite{Bekenstein2004}, namely a tensor-vector-scalar (T$e$V$e$S) theory leading to Milgrom's law in spherical symmetry as we shall see below, but {\it different} from Eq.~(\ref{eqn:poissonaqual}) in more general geometries \cite{AFZ}. Interestingly, this theory can be reformulated in various fashions, such as e.g. a pure tensor-vector theory in the matter frame \cite{Zlosnik}. We however concentrate here on the original formulation of the theory.

The tensor of the theory is the Einstein metric $g_{\alpha\beta}$ out of which is built the usual Einstein-Hilbert action, while the MONDian dynamics comes from a dynamical scalar field $\phi$ whose lagrangian density is aquadratic. All the matter fields are coupled to a physical metric involving the tensor and the scalar field of the theory \cite[eq.(21)]{Bekenstein2004}, and which must be disformally related to the Einstein metric in order to consistently enhance the deflection of photons. This needs the introduction of a dynamical normalized vector field $U_\alpha$. The total action is thus the sum of the Einstein-Hilbert action for the Einstein metric, the action of the scalar field, the action of the vector field, and the matter action involving the physical metric. Einstein-like equations are then obtained for each of these fields by varying the action w.r.t. each of them \cite{Bekenstein2004}.

The action of the scalar field $\phi$ can be written
\beq\protect\label{acttev}
S_s = -\frac{1}{2k^2l^2G} \int {\rm d}^4x \sqrt{-g} f(y)
\eeq
where $G$ is the bare gravitational constant appearing in the Einstein-Hilbert action, $g$ is the determinant of the Einstein metric, and $y = c^{-4}kl^2 h^{\alpha\beta} \partial_{\alpha} \phi \partial_{\beta} \phi$ (the metric $h^{\alpha\beta}$ being some specific combination of the
Einstein metric and the vector field \cite{Bekenstein2004}).
Here $k$ is a small dimensionless parameter (of the order of one percent) determining the strength of the scalar coupling to matter, $l$ is a length scale linked to the acceleration scale $a_0$, and $c$ is the speed of light. The function $f(y)$ is a free function driving the dynamics of the scalar field, that can be identified (to a factor) with its lagrangian density \footnote{Note that the k-essence free function $f$ is different from the free function $F$ as defined in \cite{Bekenstein2004}, note also that the scalar field $\phi$ has been chosen here to have the units of a potential, and can be seen as a usual dimensionless scalar field multiplied by $c^2$}, and that yields the MOND behaviour in galaxies when judiciously chosen. Note that for quasi-static systems (for which time derivatives are zero) such as galaxies, the scalar field is akin to k-essence scalar fields, that were notably introduced as possible {\it dark energy} fluids that could also drive inflation \cite{Damour, Chiba, Stein}. This name comes from the fact that their dynamics is dominated by their kinetic term $-f(y)$, contrary to other quintessence models in which the scalar field potential plays the crucial role.

In the Einstein frame, matter fields couple to the three
gravitational fields (tensor, vector and scalar field) through an
effective ``disformal'' metric that differs from the Einstein
metric, and which therefore, although dynamical, assumes a prior
geometric form. In that frame the above k-essence field may thus
be considered to mediate a ``fifth force'' $\vg_s = -\nabla \phi$,
in addition to the usual gravity connected to the gravitational
tensor. On the other hand, shifting our viewpoint from the
Einstein frame to the matter one, matter fields appear to couple
to some metric that assumes no prior form, but which is determined
by a non-diagonal system of coupled differential equations. In
that case the scalar field does not trivially manifest as a fifth
force. Those two points of view are however equivalent, and we
shall now work in the Einstein frame.

If one couples such a k-essence field to the matter sector in the
way it is done in T$e$V$e$S, one finds in the non-relativistic
limit, neglecting pressure, and in a static configuration, the
following modified Poisson's equation
\cite[eq.(42)]{Bekenstein2004}: \beq
\protect\label{eqn:poissonkessence} \grad.\left[\nabla \phi
f'\left(c^{-4} k l^2 |\grad \phi|^2\right)\right] = k G \rho, \eeq
where $\rho$ is the density
of the matter distribution \cite{Bekenstein2004}. The analogy with
Eq.~(\ref{eqn:poissonaqual}) is striking. For a spherical
distribution of mass $M$ one thus gets \beq
\protect\label{eqn:kessencespherique} \nabla \phi f'(c^{-4} k l^2
|\grad \phi|^2) = \frac{k G M {\bf r}}{r^3} \eeq outside the
matter. It must be stressed that many papers achieve to find
modified theories of gravity in which a test particle has an
asymptotic circular velocity of the form $v= \gamma c $ where
$\gamma$ is a pure number of the theory, and thus seem to recover
flat rotation curves of spiral galaxies \cite{Cadoni, Sobouti, Capo}. However one must not only be concerned with the existence of the plateau
velocity, but also with the whole phenomenology of rotation curves (see Sect.~II and e.g. \cite{Pers,GeSa}), including the Tully-Fisher relation
\cite{Tully, McG2}. This relation teaches us that $\gamma$ should behave as
$M^{1/4}$ (at roughly a ten percent level due to the possible
difference between the asymptotic velocity and the velocity at the
last observed radius). In
these theories it is not correct to set by hands such a value for
$\gamma$. On the contrary, k-essence theories achieve this in a
very elegant manner. It is indeed immediate to see that if $f'(y)$
behaves as $\sqrt{y}$, then the ``fifth force'' dominates the
Newton-Einstein gravity in the ultra-weak field limit, and both
the asymptotic plateau velocity and its amplitude are recovered.
K-essence theories would not have been useful if, say, the
Tully-Fisher relation were $v^p \propto M$ with $p \neq 4$. In
other words, k-essence theories are relevant toy-models in this
context because the effective (MOND) gravitational field in the
outskirts of galaxies is the square root of the Newtonian one.

\subsection{T$e$V$e$S and Milgrom's law}

To the leading order, the physical metric near a quasi-static
galaxy in T$e$V$e$S is given by the same metric as in General
Relativity, with the standard Schwarzschild potential replaced by
the total potential \beq\protect\label{total} \Phi = \Xi \phi_n +
\phi \, , \eeq where $\phi_n$ is the Newtonian potential obtained
from using the bare gravitational constant $G$, and $\Xi \sim 1$
is a dimensionless parameter which slowly evolves with the cosmic
time \cite[eq.(58)]{Bekenstein2004}. Note that this parameter can always be chosen to be precisely $\Xi=1$ at the present time. This means that, locally and at the classical level, the scalar
field $\phi$ plays exactly the role of the dark matter potential.
To the leading order, we have the following field equations in a
quasi-static system: \beq \protect\label{eqn:poisson} \grad^2
\phi_n = 4 \pi G \rho, \eeq i.e. the Poisson equation for the
Newtonian potential, and \beq \protect\label{eqn:poissonscalar}
\grad.[\mu_s(|\grad\phi| / a_0) \grad\phi] = 4 \pi G \rho, \eeq
i.e. a MOND-like equation (see Eq.~\ref{eqn:poissonaqual}) for the scalar field, obtained from
Eq.~(\ref{eqn:poissonkessence}) by defining (noticing that $y>0$
in a static configuration) \beq \protect\label{eqn:defmus}
\mu_s[(c^2/la_0)\sqrt{y/k}] \equiv 4 \pi f'(y)/k . \eeq

Note that the acceleration scale $a_0$ does not need to be precisely constant through cosmic time. However, since the variation is extremely slow, it can be considered as effectively constant during the entire time that galaxies have existed (see Sect.~III~C). From Eqs.~(\ref{total}), (\ref{eqn:poisson}), and
(\ref{eqn:poissonscalar}) we are now able to look for a Milgrom's
law for the total gravitational potential $\Phi$. Note that we
restrict ourselves to spherical symmetry so that we look for an
equation of the form \beq \protect\label{eqn:MilgromPhi}\mu(|\grad
\Phi| / a_0) \nabla \Phi = \frac{G_N M(r) {\bf r}}{r^3}, \eeq
together with the usual asymptotic MONDian conditions $\mu(x) \sim
x$ if $x \ll 1$ and $\mu(x) \sim 1$ if $x \gg 1$. The above
equation notably implies that $G_N$ is the gravitational constant
measured in Newtonian systems (and hence on Earth), which is not
required to be equal to the bare one.

For commodity we define
\beq
\vx = -\frac{\nabla \Phi}{a_0} = \frac{\vg}{a_0}, ~ x = \frac{|\grad \Phi|}{a_0} = \frac{g}{a_0}
\eeq
the MONDian gravitational force per unit mass and its modulus in units of $a_0$, and
\beq
\vs = -\frac{\nabla \phi}{a_0} = \frac{\vg_s}{a_0}, ~ s = \frac{|\grad \phi|}{a_0} = \frac{g_s}{a_0}
\eeq
the scalar field force and its modulus in units of $a_0$. Given that $a_0$ can slowly (even though not much) vary with cosmic time , the very definition of $s$ slowly varies too.

In spherical symmetry, we thus have \beq \protect\label{scalar}
\mu_s(s) s = \frac{1}{\Xi}(x-s) = \frac{|\grad \phi_n|}{a_0}. \eeq
Let us emphasize that the $\mu_s$ function can be quite
arbitrarily chosen in a k-essence theory (but see
Sect.~\ref{Sec:Cauchy}), so that the $\mu$-function that will be
derived from it may not possess the right MONDian behaviour. We
shall therefore adopt a phenomenological point of view herafter, by imposing the standard asymptotic conditions for $\mu$,
and look for the corresponding properties of $\mu_s$.

Let us define a parameter $\nu_0$ that connects the renormalized
gravitational constant measured on Earth to its bare value: $G_N=\nu_0 G$. In
spherically symmetric T$e$V$e$S, Milgrom's law can then be written
as \beq \protect\label{milgrom2} \mu(x) x =
\frac{\nu_0}{\Xi}(x-s), \eeq meaning that the $\mu$-function is defined by\beq \protect\label{mutthree} \mu
\equiv \frac{\nu_0 \mu_s}{1+\Xi \mu_s}. \eeq Note that in \cite[sect.IV~B]{Bekenstein2004} Milgrom's function was defined by \beq \protect\label{muttwo} \mut(x) x \equiv \frac{|\grad
\phi_n|}{a_0}, \eeq  neglecting the renormalization of $G$ as a first order approximation. Since it is always possible to immediately renormalize $G$ by a factor $\Xi$ such that the total potential in Eq.~(\ref{total}) is the sum of the Newtonian potential and the scalar field, it is tempting to define $\mu = \Xi \mut$. However, as we shall show hereafter, this renormalization is only valid for some special forms of the T$e$V$e$S free function. More generally the exact $\mu$-function of Eq.~(\ref{mutthree}) is related to the one of Eq.~(\ref{muttwo}) by $\mu = \nu_0 \mut$, and the fact that $\nu_0$ may differ from $\Xi$ has wide implications. 

In the rest of this section, we exhaustively examine how the free function
of T$e$V$e$S, hereafter $\mu_s(s)$ as defined in
Eq.~(\ref{eqn:defmus}), connects with the $\mu$-function of MOND,
and we insist on the importance of this famous value of $\nu_0$ inherited from the shape of $\mu_s$.

\subsection{The deep-MOND regime}

In the deep-MOND regime, we must have $\mu(x) \sim x$ for $x \ll
1$, which implies, from Eq.~(\ref{milgrom2}), that $s \ll 1$ ($s
\sim x$) and from Eq.~(\ref{scalar}) that \beq
\protect\label{eqn:DLmus}\mu_s(s) \sim \frac{s}{\nu_0} \eeq in
this regime. In order to get this behaviour in
Eq.~(\ref{eqn:defmus}) with the standard choice $f'(y) \sim
\sqrt{y/3}$ for small $y$ \cite{Bekenstein2004}, we must define
for the length scale: \beq \protect\label{l} l \equiv \frac{c^2
\sqrt{3k}}{4 \pi \nu_0 a_0}. \eeq 
Note that since $l$ is a parameter of the theory (fixed by definition), the above equation shows that $\nu_0 a_0$ must be a constant. As we will see below however, the renormalization of $G$, encoded in the parameter $\nu_0$ depends on the cosmic time, since $\nu_0$ is related to $\Xi$. This implies that the acceleration scale $a_0$ does vary with cosmic time ($a_0$ is higher at higher redshift). Let us underline that what we call $a_0$ is the quantity determining the MONDian asymptotic circular velocity for a spherical system, since in the deep-MOND regime 
test-particles feel a force given by $\sqrt{G M \nu_0 a_0}/r$ by virtue of 
Eqs.~(\ref{scalar}), (\ref{milgrom2}) and (\ref{eqn:DLmus}). This force is interpreted by an observer on Earth as a force given by $\sqrt{G_N M a_0}/r$, where $G_N=\nu_0 G$ is our local constant of gravity. However Bekenstein \cite{Bekenstein2004} estimated that the time variation of $\Xi$ is extremely mild in the matter dominated period, so that as far as fits of galaxy rotation curves are concerned, this variation of $a_0$ is expected to have a negligible influence, even though $a_0$ is expected to be somewhat different at the epoch of nucleosynthesis. 

Note also that Eq.~(\ref{l}) agrees with
\cite[eq.(62)]{Bekenstein2004} only in the very special case where
$\nu_0 = \Xi$. This special case is also very important in the
study of the Newtonian regime, as we shall see hereafter.

\subsection{The Newtonian regime: unbounded $\mu_s$-function}

The case $\nu_0 = \Xi$ is a physically interesting one, because,
{\it a priori}, it is always possible to immediately renormalize
$G$ in Eq.~(\ref{eqn:poisson}), such that $\Xi \phi_n = \Phi_N$.
Then, also replacing $G$ by $G_N=\Xi G$ in the scalar field
modified Poisson equation (Eq.~\ref{eqn:poissonscalar}) means that
the phenomenologically relevant interpolating function for the scalar field becomes 
$\Xi\mu_s$ (during a period for which the parameter $\Xi$ is constant, which is effectively the case during the entire time that galaxies have existed), and that the particular case $\nu_0 = \Xi$ thus
corresponds to no {\it a posteriori} renormalization of the
gravitational constant. Eq.~\ref{scalar}) and Eq.~(\ref{milgrom2})
can then be rewritten: 
\beq\protect\label{unbound}
x-s = \mu(x) x = \Xi\mu_s(s) s 
\eeq

The question is then whether it is actually $possible$ to have $\nu_0=\Xi$. Let us first show that we indeed have $\nu_0=\Xi$ if and only if the
$\mu_s$-function is {\it unbounded}, i.e. $\mu_s \rightarrow
\infty$. Indeed, if $\nu_0 = \Xi$, then, from the first equality in
Eq.~(\ref{unbound}), we can write $\mu$ as $1-s/x$ where we must
have $s \ll x$ for the Newtonian regime $x \gg 1$, which
proves from the second equality in Eq.~(\ref{unbound}) $s = x/(1+\Xi\mu_s)$
that $\mu_s$ must diverge with $s$. Respectively, if $\mu_s(s)$
diverges for some value $s=s_0$, it is straightforward from
Eq.~(\ref{mutthree}) that $\mu \rightarrow \nu_0/\Xi$ and therefore
that $\nu_0 = \Xi$ by definition of $\mu$.

Then let us emphasize that Eq.~(\ref{unbound}) actually corresponds to the implicit assumptions in \cite{ZF06}, where it was shown that, without any {\it a posteriori} renormalization of the gravitational constant, it was impossible to find a T$e$V$e$S free function $\mu_s(s)$ corresponding to the standard $\mu$-function of MOND (Eq.~\ref{standard}).
This particular result can actually be generalized through the following
{\it theorem:}

\noindent When $\nu_0 = \Xi$, a MOND $\mu$-function can be obtained from a T$e$V$e$S free function {\it only if} $n \leq 1$ in its asymptotic expansion $\mu(x) \sim 1 - A/x^n$ (where $A$ is some positive constant).

\noindent The {\it proof} is quite straightforward:

\noindent Since $\nu_0 = \Xi$, we have from Eq.~(\ref{milgrom2}) (or from the first equality in Eq.~\ref{unbound})
\beq \protect\label{invtheorem}
s = x [1-\mu(x)] \, .
\eeq
Let us assume that $n>1$, then for $x \rightarrow \infty$ we would
have $s \sim A/x^{n-1} \rightarrow 0$. But since $s \ll 1$ is already the deep-MOND regime (see Sect.~III~C), it cannot be also the Newtonian one $\square$

What it means is that, as can be easily checked \cite{ZF06}, the
$\mu_s$ function would be multivalued. This is however not
acceptable since $\mu_s$ must define unambiguously the dynamics of
the scalar field. No unbounded $\mu_s$-function can thus reproduce
the standard $\mu$-function (Eq.~\ref{standard}), for the reason
that the latter admits the following asymptotic expansion \beq
\mu(x) \sim 1 - \frac{1}{2 x^2}, \eeq and by virtue of the
previous theorem (here $n=2$).

On the other hand, if $n < 1$, this is not a problem since $s$ would diverge jointly with $x$, while if $n=1$, $s$ would asymptote to a finite positive value, and the function $s(x)$ from Eq.~(\ref{invtheorem}) could still be strictly increasing, and thus be invertible. E.g., the simple $\mu$-function of MOND, (Eq.~\ref{simple}),
has $n=1$ for its asymptotic expansion. Its corresponding function $s(x)$ from Eq.~(\ref{invtheorem}) is
\beq \protect\label{scalarsimple}
s(x) = \frac{x}{1+x},
\eeq
leading to the unbounded $\mu_s$ function
\beq \protect\label{simplescalar}
\Xi \mu_s(s) = \frac{s}{1-s}.
\eeq
This corresponds to the $n=0$ case of the free function proposed by \cite[eq.(13)]{ZF06} for the local value $\Xi=1$. The $\mu$-functions tested against observations in Sect.~II also have an unbounded $\mu_s$ counterpart in T$e$V$e$S. If we want their corresponding k-essence free functions to be expressed as pure functions of the constants in the action of T$e$V$e$S (see Sect.~III~A), independently of cosmic variations of $\Xi$, we can write the $\alpha$-family of Eq.~(\ref{eqalpha}) as:
\beq\protect\label{alphafam}
\mu_s(s) = \frac{s}{\Xi - \alpha s}.
\eeq

Another interesting aspect of these unbounded $\mu_s$ functions (or of the special case $\nu_0 = \Xi$) is that they lead to an analogy with electromagnetism that might be helpful to grasp the feeling of the to-be-understood physical meaning of the scalar field of T$e$V$e$S. Indeed, with the definition of Eq.~(\ref{mutthree}), one can then write the Newtonian gravity generated by a baryonic point mass $M$ as:
\beq
\frac{G_N M}{r^2} = g_N = \mu \times (g_N + g_s) = \frac{\mu}{1-\mu} \times g_s.
\eeq
In general, the conservative electric field $\vE$ is given by
\beq
\vE = \vd + \vp, ~  {\rm with} ~  \vd \equiv \frac{\vD}{\epsilon_0}, ~ \vp \equiv -\frac{ \vP}{\epsilon_0 },
\eeq
where $\epsilon_0$ is the absolute permissivity of the vaccum, $\vD$ is the
electric displacement vector, and $\vP$ is the polarization vector.
If we consider a dielectric medium of relative permissivity $\mu$
surrounding a free charge $-Q$, we have
\beq
-\frac{Q}{r^2} = d = \mu \times (d + p) = \frac{\mu}{1-\mu} \times p.
\eeq
The analogy with T$e$V$e$S is striking, the free charge $-Q$ playing the role of a baryonic point mass $M$ (both creating a $1/r^2$ force in the absence of the ``dielectric'').
The scalar field `` fifth" force $\vg_s$ plays the role of the polarization field $\vp$ and the total gravitional acceleration $\vg$
plays the role of the conservative electric field $\vE$. The electromagnetic dielectric
permissivity $\mu$ is often constant for a uniform medium, depending on the
microscopic structure of the medium, but can vary spatially in the presence
of a non-uniform field. Interestingly the polarization field in a
polarizable dielectric medium has also an microscopic couterpart:
the randomly oriented dipoles in the medium are
lined up to form a macroscopic dipole (of amplitude $P$ per unit volume)
when the electric field $\vE$ is applied.
Could this hint towards a quantum origin
for a T$e$V$e$S-like field theory, with the vacuum playing
the role of the dielectric medium, and the scalar field force arising
from the gravitational effect of the baryonic matter? This is all too
speculative for the time being, and this analogy should of course break down at the cosmological level in T$e$V$e$S because of the role played by the vector field. This is pretty much beyond the scope of this
paper, but we hope the analogy may help the reader to grasp the feeling of one possible concrete physical meaning of the scalar field.

What is more, it is actually striking that, for the best-fit $\alpha=1$ function of Sect.~II and at $\Xi=1$, the $\mu$-factor of Eq.~(1), $is$ nothing else than the scalar field strength itself in units of $a_0$, i.e.
\beq\protect\label{equivalence}
\mu(x)=s(x)=\frac{|\grad \phi|}{a_0},
\eeq
from Eq.~(\ref{simple}) and Eq.~(\ref{scalarsimple}).
This may acquire a great theoritical significance in any future T$e$V$e$S-like theory where the $\mu$-function could be viewed as a dielectric-like factor. The fact that what could be interpreted as a ``digravitational" relative permissivity of space-time could simply be given by the modulus of the scalar field gradient is very appealing and may pave the way towards a well-motivated relativistic theory of MOND. We stress that this relation only holds locally in T$e$V$e$S because of the time-evolution of $\Xi$ (although this parameter is effectively constant during the entire matter dominated period), but that this factor comes from the vector field in T$e$V$e$S, and could thus perhaps be avoided in somewhat different relativistic theories of MOND \cite{Blanchet2, JPBGEF}.

\subsection{The Newtonian regime: bounded $\mu_s$-function}

We are now going to show that the above theorem constraining the MOND $\mu$-function in the context of unbounded T$e$V$e$S $\mu_s$-functions breaks down when $\nu_0$ differs from $\Xi$ (something which was not realized in \cite{ZF06}).

If $\nu_0 \neq \Xi$, then, using the condition $\mu(x)
\rightarrow 1$ for $x \rightarrow \infty$ (i.e. the MOND condition
to recover Newtonian dynamics in the high acceleration regime), we
find from Eq.~(\ref{milgrom2}) and Eq.~(\ref{scalar}) that $s \gg 1$
and that the $\mu_s$ function is $bounded$, i.e. asymptotes to a finite value which
reads \beq \mu_0 = \frac{1}{\nu_0-\Xi}. \eeq Note that $\mu_s$
must be strictly positive (see Sect.~\ref{Sec:Cauchy}) so that
$\nu_0
> \Xi$. Respectively, it is straightforward from Eq.~(\ref{mutthree})
that if $\mu_s(s) \rightarrow \mu_0$ for $s \rightarrow \infty$,
then we have that $\mu \rightarrow \nu_0\mu_0/(1+\Xi\mu_0)=1$,
meaning that $\nu_0 = \Xi + 1/\mu_0$.

The above theorem for unbounded functions then breaks down because, in order to get a
$\mu$-function with an asymptotic expansion $\mu(x) \sim 1 -
A/x^n$, we need a scalar field strength such that \beq s \sim
\frac{x}{1+\Xi\mu_0} + \frac{A\Xi\mu_0}{1+\Xi\mu_0} x^{-n+1}, \eeq
meaning that, when $x$ is large, $s$ does not tend to zero, and
$\mu_s(s)$ is not necessarily multivalued when $n>1$.

It is actually possible to find such bounded $\mu_s$
functions that give rise to the standard $\mu$-function of
Eq.~(\ref{standard}). There is in fact a one-parameter family of
such functions (labelled by the value $\mu_s(\infty)=\mu_0$, or
equivalently by $\nu_0$). When
inserting the standard $\mu$-function of Eq.~(\ref{standard}) in
Eq.~(\ref{milgrom2}), one must make sure that the function \beq \label{invertible} s =
x \left[ 1 - \frac{\Xi}{\nu_0} \mu(x) \right] \eeq is strictly
increasing, and thus invertible. This means that me must have
$\nu_0/\Xi > 4/3 \sqrt{2/3}$.

However, such bounded $\mu_s$-functions lead to a non-trivial renormalization of the gravitational constant, and their relation with the $\mu$-function of MOND is much less elegant than in the case of unbounded functions, leading to a breakdown of the analogy with electromagnetism. Indeed, in the case of unbounded functions, the relevant function for the scalar field could be considered to be $\Xi\mu_s$, and one needed $\Xi\mu_s = s$ for small $s$ to recover $\mu(x)=x$ for small $x$.
With the definition of $l$ in \cite[eq.(62)]{Bekenstein2004}, this is exactly the behaviour of $\mu_s$ for Bekenstein's toy model at small $s$. However, since Bekenstein's $\mu_s$ is a {\it bounded} function, its corresponding MOND $\mu$-function is proportional to $\nu_0x/\Xi$ instead of $x$ in the weak regime, because of the renormalization of $G$. This means that, in order to recover the Tully-Fisher law in weak gravity, $l$ must be redefined as in Eq.~(\ref{l}). But the $\mu$-function of Eq.~(\ref{bekmu}) is then only a rough approximation of Bekenstein's toy model, even in the intermediate regime, because of the renormalization of $G$ that makes the $\mu$-function asymptote more quickly to $\mu=1$ than Eq.~(\ref{bekmu}). On the other hand, note that there exists a one-parameter family of bounded $\mu_s$-functions $precisely$ yielding Eq.~(\ref{bekmu}), but different from Bekenstein's free function, while for a given $\Xi$ the {\it unbounded} function yielding Eq.~(\ref{bekmu}) is simply $\Xi \mu_s= s$.

\subsection{Non-Spherical systems}

The relations between $\mu$ and $\mu_s$ we derived above are of course only valid in spherical symmetry. Out of spherical symmetry, Milgrom's law in Eq.~(1) is not exact for any MOND-like gravity theory: the Newtonian force, the classical MOND force, and the scalar field force are no longer parallel. The curl field obtained when solving for the full classical MOND potential $\Phi$ in Eq.~(\ref{eqn:poissonaqual}) will be different from the one obtained when solving the equation
for the scalar field $\phi$ (Eq.~\ref{eqn:poissonscalar}) (see e.g. \cite{AFZ}). Note that, for an unbounded $\mu_s$-function ($\nu_0 = \Xi$), the latter can be rigorously remoulded into the following expression
\begin{equation} \protect\label{nonsymteves}
\vg = \vg_{N} + \vg_s, \qquad \grad \cdot \left[ \Xi \mu_s \vg_s - \vg_{N} \right] = 0.
\end{equation}
Here the total gravity $\vg$ consists of a baryonic Newtonian gravity
$\vg_{N}$ (calculated from the Newtonian Poisson equation with the Newtonian gravitational constant $G_N = \Xi G$) plus $\vg_s$ as an extra or boosted gravity (the space part of the four vector gradient of the scalar field $\phi$ of T$e$V$e$S, simply playing the role of dark matter gravity). The amount of boost is specified by $\Xi \mu_s$, which takes the following dual meaning: (1) the ratio of baryonic
gravity w.r.t. extra-gravity, or (2) the ratio of {\it effective} inertial mass $m_{\rm eff}=m\mu$ (not to be confused with the actual inertial mass, unmodified in a modified $gravity$ theory) to the effective loss of inertial mass $m(1-\mu)$, with $\mu$ defined as in Eq.~(\ref{mutthree}).

\begin{figure}
   \centering
   \includegraphics[width=8cm]{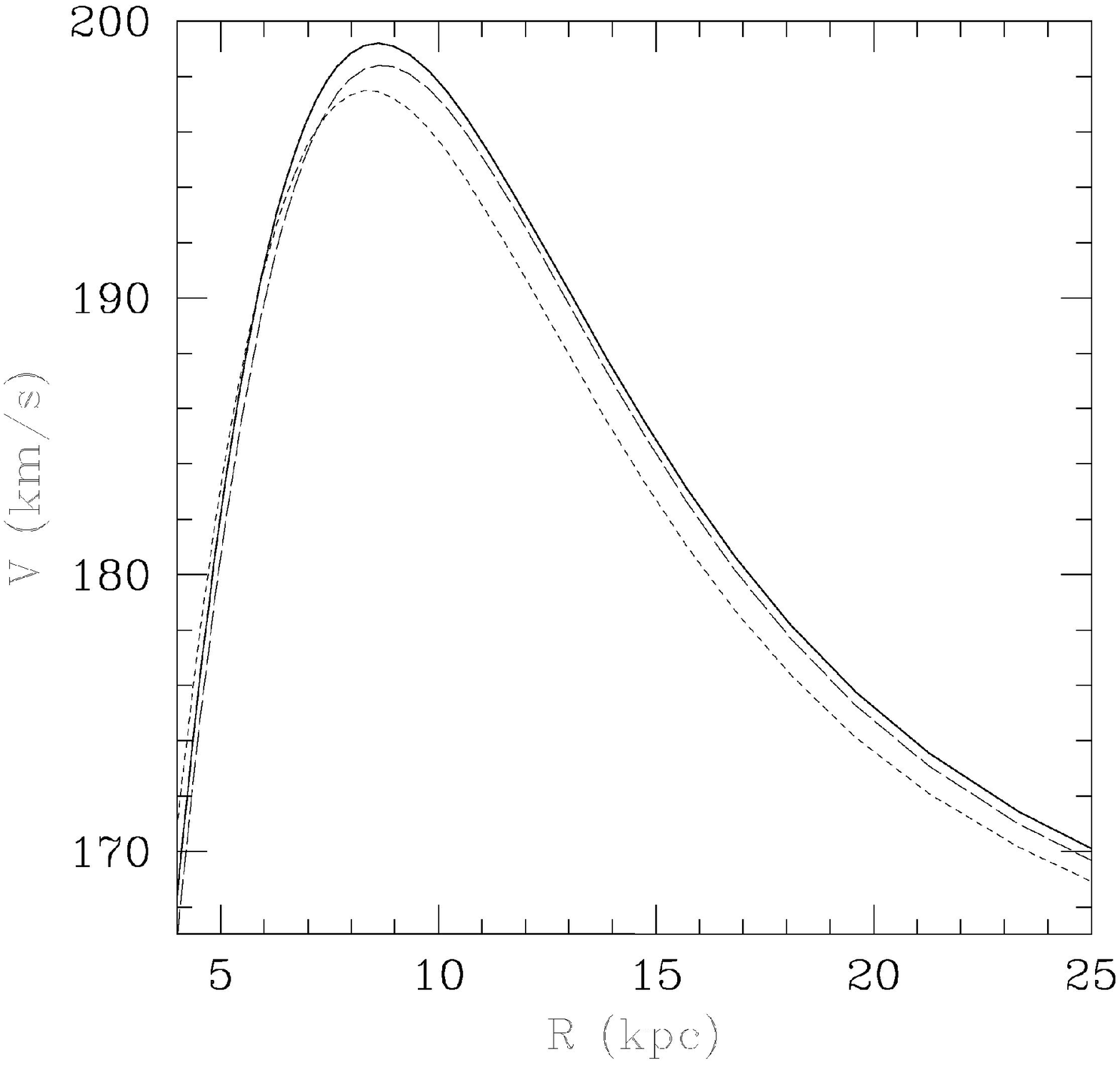}
   \includegraphics[width=8cm]{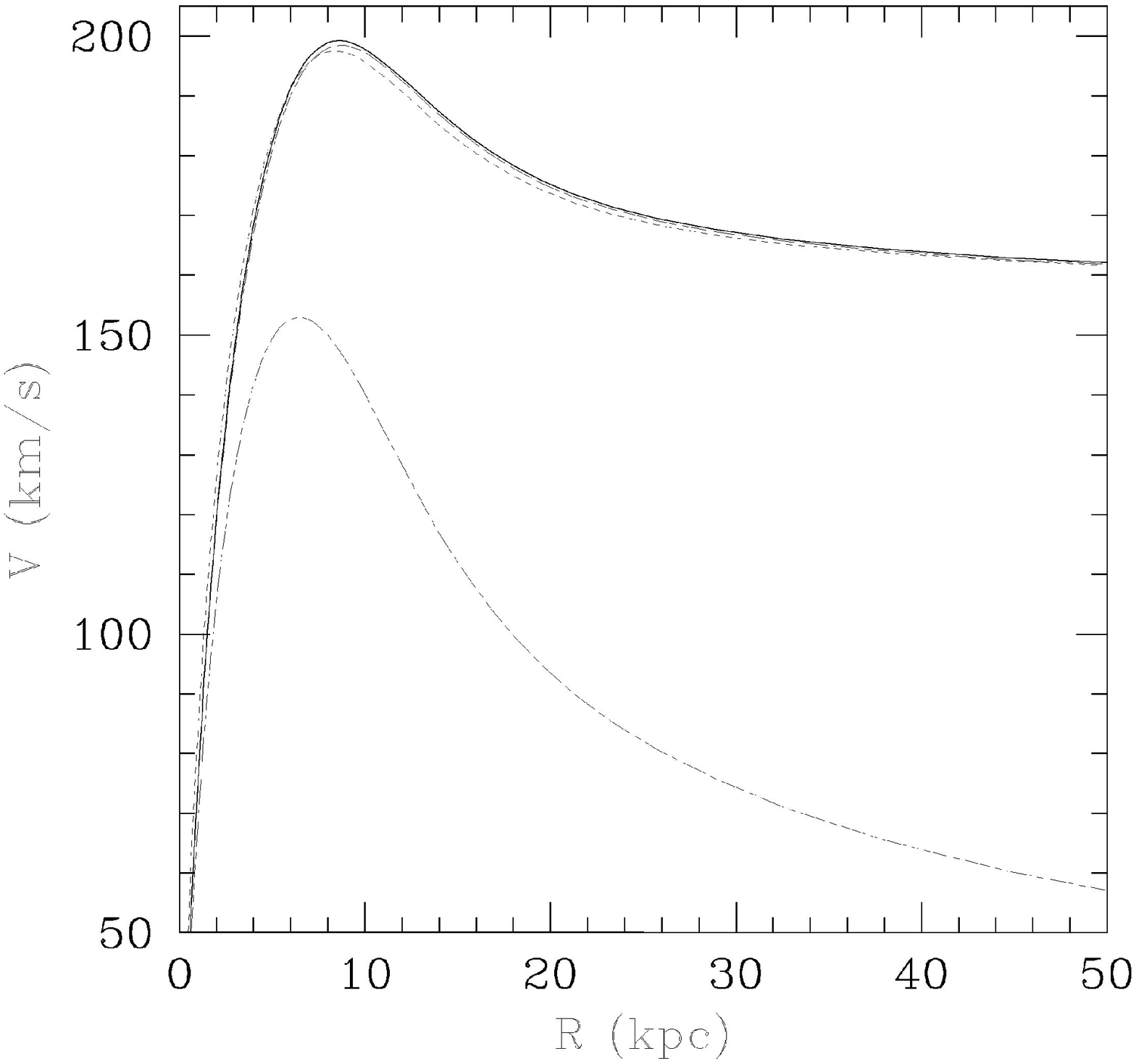}
      \caption{Bottom: exact Newtonian (dot-dashed line), classical
      MOND (dashed line), and T$e$V$e$S (full line) equatorial-plane rotation curves for a disk galaxy model with density
      distribution given in equation (\ref{eqexpd}). The dotted line is
      the corresponding rotation
      curve approximated using Eq.(1). The top panel is a zoom-in of the upper left corner of the bottom panel.}
\label{figrcfull}
   \end{figure}

Here, for the first time in the literature, we compare the different computed rotation curves for the same disk galaxy using the prescriptions of Eq.~(1) (Milgrom's law), Eq.~(\ref{eqn:poissonaqual}) (classical MOND), and Eq.~(\ref{eqn:poissonscalar}) (or equivalently Eq.~\ref{nonsymteves}, i.e. T$e$V$e$S). We consider an exponential disk model typical of a high surface brightness galaxy,
characterized by a baryonic density distribution
\begin{equation}
\rho_b(R,z) = \frac{M_b}{4 \pi {R_0}^2 z_0} {\exp\left(-\frac{R}{R_0}\right) (\cosh \frac{z}{z_0})^{-2}},
\label{eqexpd}
\end{equation}
with total mass $M_b=4.2\times 10^{10}\Msun$, and characteristic scale
lengths $R_0=3 \kpc$ and $z_0=0.1 \kpc$, where $R$ and $z$ are the
cylindrical coordinates.

We can compute the exact force field, by solving
Eq.~(\ref{eqn:poissonaqual}) using the numerical MOND potential solver
developed by \cite{CLN}. This solver was
originally tested in \cite{CLN} for the standard
$\mu$-function of Eq.~(\ref{standard}), but we verified it
works as well in solving Eq.~(\ref{eqn:poissonaqual}) for other $\mu$-functions such as Eq.~(\ref{simple}) and Eq.~(\ref{bekmu}), or in solving the scalar field Poisson equation for a $\mu_s$-function such as Eq.~(\ref{simplescalar}).

For this model we compute {\it exactly}
four different equatorial-plane rotation curves, which are plotted in
Fig.~\ref{figrcfull}: the Newtonian rotation curve (dot-dashed), the classical MOND rotation curve for the simple $\mu$-function of Eq.~(\ref{simple})(dashed), and the T$e$V$e$S
rotation curve for the $\mu_s$-function of Eq.~(\ref{simplescalar}) (full line; we assumed for this plot $a_0=1.2\times 10^{-8}$ cm s$^{-2}$).  The MOND/T$e$V$e$S rotation curves deviate significantly from the Newtonian one even at small radii, which is characteristic of the simple function (see Sect.~II). Fig.~\ref{figrcfull} also
plots as a dotted line the approximated rotation curve calculated from Eq.~(1) with the simple $\mu$-function. From the bottom panel of Fig.~\ref{figrcfull} it is apparent how for an exponential disk model the exact T$e$V$e$S rotation curve, the exact classical MOND rotation curve, and the corresponding approximated rotation curve are practically
indistinguishable from an observational point of view. Only a zoom-in
at intermediate radii (Fig.~\ref{figrcfull}, top) allows to appreciate
the difference between the three modified-gravity curves: the T$e$V$e$S
curve is systematically slightly higher (by less than 0.5\%) than the classical MOND
one, while the approximated curve using Milgrom's law slightly overestimates the velocity (by only a few tenth of percent) for $R <~2R_0$, and slightly underestimates it (by less than 1\%) for $R >~ 2R_0$. Note that if we consider $R=25$~kpc as the typical last observed radius in this high surface brightness galaxy, the difference between the circular velocity at that point and the asymptotic one ($\sim$ 5-10~km/s) is in any case higher than the difference between the velocities predicted at that point from Milgrom's law and from T$e$V$e$S due to the non-sphericity of the system ($\sim$~1-2~km/s). 

\section{Cosmology}

\subsection{\protect\label{Sec:Cauchy}The Cauchy problem}

We have seen that the asymptotic condition $\mu(x) \sim x $ when $x \ll 1$ implies that $\mu_s(s)$ must behave linearly with $s$ when $s$ is small (Eq.~\ref{eqn:DLmus}), and notably that we must have $\mu_s(0)=0$. This $s=0$ signals the transition between local physics and cosmology in T$e$V$e$S \cite{Bekenstein2004}.

Let us however emphasize that a k-essence theory such as T$e$V$e$S can exhibit
superluminal propagations whenever $f''(y) >0$ \cite{JPB06}. Although it does
not threaten causality \cite{JPB06}, one has to check that the
Cauchy problem is still well-posed for the field equations. Within the choice of signature
$(-,+,+,+)$ of local Lorentzian metrics, it has been shown \cite{
JPB06,Susskind,Rendall} that it requires the otherwise free
function
$f$ to satisfy the following properties, $\forall y$: \bey f'(y) > 0 \\
f'(y) + 2 y f''(y) >0. \eey
In terms of the scalar field strength $s \equiv (c^2/la_0)\sqrt{y/k}$ (when $y \geq 0$, for a quasi-static galaxy), and of the auxiliary function $\mu_s(s) \equiv 4 \pi f'(y)/k$ , these conditions read, $\forall s \geq 0$: \bey \mu_s(s) > 0 \\
\mu_s(s)+ s \mu'_s(s) >0. \eey
where the prime denotes the
derivative with respect to $s$. Note that, throughout Sect.~III, although we have shown that it was sometimes impossible to recover a given $\mu$-function from any single-valued unbounded $\mu_s$-function, we implicitly assumed that for a given $\mu_s$ it was $always$ possible to recover a spherically symmetric formulation ``\`a la Milgrom" of local physics of T$e$V$e$S, i.e. an equation like Eq.~(\ref{eqn:MilgromPhi}). This means that when inserting a given $\mu_s$ in Eq.~(\ref{scalar}), the inversion of $x(s)$ was assumed to be $always$ possible (i.e. $x(s)$ is strictly increasing). Making use of $\Xi >0$, this is actually $guaranteed$ by the second inequality above, meaning that $s \mu_s(s)$ must be a strictly increasing function of $s$.

Now, concerning the transition between local physics and cosmology in T$e$V$e$S, the asymptotic behaviour of the
$\mu$-function can therefore \textit{not} be implemented by a
consistent k-essence theory, since  $\mu_s(0)=0$ would violate both inequalities at $s=0$. As
explained in \cite{JPB06,JPBGEF},  all present day k-essence-like
theories of MOND such as T$e$V$e$S predict the
existence of a singular surface around each galaxies on which the
scalar degree of freedom does not propagate, and can therefore not
provide a consistent picture of collapsed matter embedded into a
cosmological background.

A simple solution consists in assuming a modified asymptotic
behaviour of the $\mu$-function, namely of the form \beq \mu(x)
\sim \varepsilon_0 + x \eeq if $x \ll 1$. In that case there is a
return to a Newtonian behaviour at a very low acceleration scale $x
\ll \varepsilon_0$, and rotation curves of galaxies are only
approximatively flat until the galactocentric radius \beq R \sim
\frac{1}{\varepsilon_0}\sqrt{\frac{G_N M} {a_0}}.
\eeq One must thus have $\varepsilon_0 \ll 1$. A 
similar phenomenology also arises in a slightly different context in \cite{Sanderseps}. Present data on galaxy rotation curves \cite{Gen3741} suggest that $1/\varepsilon_0$ must be at least of the order of $10$. Note that $\varepsilon_0$ is a quantity which slowly evolves with cosmic time (such as $\nu_0$ and $a_0$). In order to have such a behaviour for $\mu$, one must have in the deep-MOND regime $\mu_s \sim s/\nu_0 + \varepsilon$ for small $s$, and we then have $\varepsilon_0 = \nu_0 \varepsilon$ (where $\varepsilon$ is a constant).

Note that, in the negative range of values of $y$ corresponding to the time-like sector of the expanding Universe, $\mu_s$ can still be defined as a function of $y$ as in Eq.~(\ref{eqn:defmus}). It can also be viewed as a function of $s$ if we extend the definition of s to
\beq\protect\label{defis}
s \equiv {\rm sign}(y) \frac{c^2}{l a_0} \sqrt{\frac{|y|}{k}}.
\eeq
Then the Cauchy conditions read, $\forall s \leq 0$:
\bey \mu_s(s) > 0 \\
\mu_s(s)- s \mu'_s(s) >0. \eey

\subsection{The mirror-image function}

To be able to describe both galaxies and cosmology,  the free function of T$e$V$e$S must thus be able to deal also with non-stationary systems such as the Universe itself. We describe hereafter how this can be achieved, following \cite{ZF06}.

Bekenstein's original proposal \cite{Bekenstein2004} was to construct the scalar field action of Eq.~(\ref{acttev}) whith a lagrangian density given as a one-to-one function of $\mu_s$ viewed as an auxiliary non-dynamical scalar field, i.e. $f(y) = {\cal F}(\mu_s)$.

Such one-to-one construction had the drawback that the lagrangian density of the scalar field necessarily had unphysical ``gaps", i.e. that a sector was reserved for space-like systems (e.g. from dwarf galaxies to the solar system in the range $0<\mu_s<\mu_0$ for $y>0$) and a disconnected sector was reserved for time-like systems (e.g., the expanding Universe in the range $\mu_s>2\mu_0$ for $y<0$). While viable mathematically, such a disconnected Universe would not permit galaxies to collapse continuously out of the Hubble expansion.

In an effort to re-connect galaxies with the expanding Universe in T$e$V$e$S, it was proposed in \cite{ZF06} to construct the lagrangian density as a one-to-one {\it continuous} function of $y$, rather than a function of $\mu_s$, thus not considering $\mu_s$ as an auxiliary non-dynamical scalar field anymore. This allows for a smooth transition from the edge of galaxies where $y \sim 0$ to the Hubble expansion. It was also suggested in \cite{ZF06} to extrapolate the lagrangian density for galaxies into the cosmological regime by simple mirror-imaging, to minimize any
fine-tuning in T$e$V$e$S, i.e.
\beq
\mu_s(s) = \mu_s(-s)
\eeq
with $s$ defined as in Eq.~(\ref{defis}), meaning that the time-like sector is a simple mirror-image of the space-like sector.

In other words, it means that, in order to produce a consistent cosmological picture, we suggest to extend the $\alpha$-family of Eq.~(\ref{alphafam}) to the time-like sector by using the following k-essence function $f(y)$ in the action of the scalar field (Eq.~\ref{acttev}):
\beq
f(y) = f(0) + \int_0^y \frac{k \mu_s}{4 \pi} {\rm d}y
\eeq
with
\beq
\mu_s = \varepsilon + \frac{\sqrt{|y|}}{A_0 - \alpha \sqrt{|y|}}, ~ 0< \varepsilon < 0.1, ~ 0
\leq \alpha \leq 1,
\eeq
where
$A_0 = \sqrt{3} k/4 \pi$.

It would thus be of great interest to compute the CMB anisotropies and matter power spectrum \cite{Skordis, Dodelson} using this action for the scalar field allowing for a smooth transition from the edge of galaxies to the Hubble expansion. Note also that the integration constant $f(0)$ can play the role of the cosmological constant \cite{Hao} ($f(0) = 0$ corresponding to no cosmological constant), but that even a no-cosmological constant model could drive late-time acceleration \cite{Zcosmo, Diaz}, which is not surprising since k-essence scalar fields were first introduced to address the dark energy problem. This T$e$V$e$S cosmological model is thus worth exploring since it could have the ability to resolve many problems at the same time. However, the $\alpha$-family of $\mu$-functions does not work very well on galaxy cluster scales if $\vg_N$ is the gravity generated by baryons alone in Eq.~(\ref{nonsymteves}). This problem of the MOND-phenomenology has been known for a while \cite{Aguirre}, and it was then suggested \cite{San2003} that ordinary neutrinos of 2~eV could be included in the ``known matter-total gravity" relation. Including $\vg_N$ as the gravity of both baryons and neutrinos, it was found in \cite{ASZF} that a reasonable fit to the lensing of the now famous bullet cluster \cite{Clowe} is possible if neutrinos have a 2~eV mass (a mass also invoked in current MOND fits of the CMB anisotropies \cite{Skordis, McGCMB}), while the unusually high encounter speed of the bullet cluster \cite{bulvel} could be reproduced since the modified gravity potentials are deeper than the usual CDM ones. This cosmological framework was called the $\mu$HDM paradigm in \cite{ASZF} by contrast with the usual $\Lambda$CDM one. These neutrinos would not change the gravity on galaxy scales due to their low densities \cite{ASZF}, hence would not affect the goodness of the rotation curves fits presented in Sect.~II. However, for elliptical galaxies at the very center of galaxy clusters where the neutrino density is the highest, the velocity dispersion profiles could be affected by these neutrinos, thus explaining the poor purely baryonic fits for NGC~1399 when using Eq.~(\ref{simple}) \cite{Richtler} (even though the external field effect in this galaxy, due to the gravity of the cluster itself, might render the fits difficult to achieve in practice even when including neutrinos).  

\section{Conclusions}

Throughout this paper we have studied from a theoretical point of view the observationally motivated baryon-gravity relation in galaxies \cite{Don, McG, McGa}. This relation is completely encapsulated within the $\mu$-function (see Eq.~1) of Modified Newtonian Dynamics \cite{Milgrom83}, each distinct function defining a distinct baryon-gravity relation. This relation might be empirical or fundamental, but the present-day success of the relation with the sole constraints of the MOND paradigm, i.e. $\mu(x)=1$ at $x\gg 1$ and $\mu(x)=x$ at $x\ll 1$ with $x \sim g/cH_0$, points at first sight to a modification of gravity in the ultra-weak field limit. Even though the most important aspect of MOND remains this asymptotic behaviour, the study of the transition between the Newtonian regime and the deep-MOND regime, characterized by the shape of the $\mu$-function, may also be important in the quest for a possible fundamental theory underpinning the MOND prescription. We have thus investigated the observational constraints upon the $\mu$-function in that transition zone, as well as the theoretical constraints upon it in the framework of the relativistic formulation of MOND dubbed T$e$V$e$S \cite{Bekenstein2004}. Note that the latter are only valid if MOND really has its basis in a multifield theory like T$e$V$e$S, which is of course far from guaranteed. 

\noindent We have shown that:

\noindent {\bf i)} Observationally (from the quality of galaxy rotation curves and the current accuracy of population synthesis models), one cannot distinguish between the ``simple" (Eq.~\ref{simple}) and the ``standard" (Eq.~\ref{standard}) form of the $\mu$-function, but the transition from Newton to MOND cannot be much more gradual than that implied by the simple function (Sect.~II).

\noindent {\bf ii)} The simple form of $\mu$ does require generally lower mass-to-light ratios for the stellar components of spiral galaxies than does the standard form. This could, combined with accurate population synthesis models, eventually distinguish between them (Sect.~II).

\noindent {\bf iii)} A particular class of T$e$V$e$S free functions (the unbounded $\mu_s$ functions in Eq.~\ref{eqn:poissonscalar} and Eq.~\ref{eqn:defmus}) correspond to a trivial renormalization of the gravitational constant $G_N = \Xi G$ where $\Xi$ is a parameter of the theory, and are easily linked to the $\mu$-function of MOND. However, this $\mu$-function must have $n \leq 1$ in its asymptotic expansion $\mu(x) \sim 1 - A/x^n$ (where $A$ is some positive constant), meaning that the standard form of $\mu$ (Eq.~\ref{standard}) cannot be recovered from a single-valued unbounded $\mu_s$ (Sect.~III~D).

\noindent {\bf iv)} Renormalizing the gravitational constant as $G_N = \nu_0 G$ with $\nu_0 > \Xi$ for bounded $\mu_s$ functions $allows$ to recover $\mu$-functions with a steeper asymptotic expansion. The standard $\mu$ of Eq.~(\ref{standard}) can be recovered if $\nu_0$ is such that Eq.~(\ref{invertible}) is invertible (Sect.~III~E).

\noindent {\bf v)} Observationally (from galaxy rotation curves) one cannot distinguish between Milgrom's law \cite{Milgrom83}, classical MOND \cite{BM84} and T$e$V$e$S \cite{Bekenstein2004}. Using the code of \cite{CLN} we have shown that the difference between the circular velocity at the last observed radius in a typical spiral galaxy and the asymptotic circular velocity is higher than the difference between the velocities predicted at that point from classical MOND and from T$e$V$e$S due to the non-sphericity of the system (Sect.~III~F).

\noindent {\bf vi)} Some modification of the $\mu$-function is necessary to make cosmology consistent with quasi-static mass distributions in T$e$V$e$S, which should lead to a return to $1/r^2$ attraction at very low accelerations in galaxies (Sect.~IV~A).

\noindent {\bf vii)} By extrapolating the scalar field lagrangian density for galaxies into the cosmological regime by simple mirror-imaging in T$e$V$e$S, one minimizes the fine-tuning of the theory while avoiding that the sector reserved for space-like systems such as galaxies be disconnected from the sector reserved for time-like systems such as the Universe itself (Sect.~IV~B).

Let us finally note that the simple form of $\mu$ (Eq.~\ref{simple}), yielding excellent fits to galaxy rotation curves and not needing a non-trivial renormalization of the gravitational constant in T$e$V$e$S, has to be distorded in the limit of high accelerations to be in accordance with observed planetary motions in the inner solar system \cite{sanderssolar}. This could be achieved for an unbounded $\mu_s$ by e.g. somehow making the parameter $\alpha$ dynamical in Eq.~(\ref{alphafam}), going smoothly from 1 in galaxies towards 20 at the orbit of Mercury. More prosaically, it can be achieved by making the $\mu_s$ function bounded as in the $n=1$ case of \cite[eq.(13)]{ZF06}.

\begin{acknowledgments}
We are grateful to Carlo Nipoti for his kind help with the computation of rotation curves in non-spherical geometries. We thank Kor Begeman for kindly providing us his rotation curve data, as well as Stacy McGaugh, Jacob Bekenstein, Pedro Ferreira, Garry Angus and Tom Zlosnik for insightful comments. We also thank the referee for his careful reading of the manuscript and his comments that have greatly improved the clarity of the paper. BF is a Research Associate of the FNRS.
\end{acknowledgments}

\end{document}